\documentstyle[prl,aps]{revtex}

\twocolumn

\begin{document}

\title{Lifetimes of image-potential states on copper surfaces} 
\author{E. V. Chulkov$^1$, I. Sarr\'\i a$^2$, V. M. Silkin$^1$, J. M. Pitarke$^2$,  and 
P. M. Echenique$^{1,3}$} 
\address{$^1$Materialen Fisika Saila, Kimika Fakultatea, Euskal Herriko
Unibertsitatea,\\  1072 Posta kutxatila, 20080 Donostia, Basque Country, Spain\\
$^2$Materia Kondentsatuaren Fisika Saila, Zientzi Fakultatea, Euskal Herriko  Unibertsitatea,
\\ 644 Posta kutxatila, 48080 Bilbo, Basque Country, Spain\\
$^3$Unidad Asociada al Instituto de Ciencia de Materiales, C.S.I.C., Cantoblanco, 28049 Madrid,
Spain}
\date\today
\maketitle

\begin{abstract}
The lifetime of image states, which represent a key quantity to probe the coupling of
surface electronic states with the solid substrate, have been recently determined for quantum
numbers $n\le 6$ on Cu(100) by using time-resolved two-photon photoemission in combination with the
coherent excitation of several states (U. H\"ofer {\it et al}, Science {\bf 277}, 1480 (1997)). We
here report theoretical investigations of the lifetime of image states on copper
surfaces. We evaluate the lifetimes from the knowledge of the self-energy of the excited
quasiparticle, which we compute within the GW approximation of many-body theory. Single-particle
wave functions are obtained by solving the Schr\"odinger equation with a realistic one-dimensional
model potential, and the screened interaction is evaluated in the random-phase approximation
(RPA). Our results are in good agreement with the experimentally determined decay
times.
\end{abstract}


\narrowtext

In a variety of metal surfaces, the bulk band structure projected onto the surface presents a band
gap near the vacuum level. Thus, given an electron located outside such a metal
surface, it may be trapped in the vacuum well produced by the self-interaction of the electron with
the polarization charge it induces in the surface\cite{Echenique2,Osgood}. Far from the
surface, into the vacuum, this potential well approaches the long-range classical image potential,
and the resulting quantized electronic states form a Rydberg-like series which converges towards the
vacuum energy and is, in principle, resolvable\cite{Echenique2}. These so-called image states were
first identified experimentally\cite{Dose,Himpsel} by inverse photoemission\cite{Pendry,Smith1}, and
the first high-resolution measurements of image states were performed by the use of two-photon
photoemission (2PPE)\cite{Giesen,Padowitz,Fauster,Harris}. Combined with the use of ultrafast lasers,
2PPE has provided a powerful technique to directly probe, on a femtosecond time scale, electron
dynamics of excited electrons in metals\cite{Aeschlimann,Hertel} and, in particular, image
states\cite{Hertel,Wolf,Hofer2,Hofer}.

Besides their well-defined physical properties,
image-potential induced states are of general interest in surface science, because of their
applicability to other areas of condensed-matter physics, providing, in particular, a very simple
model to investigate the coupling of surface electronic states with the solid substrate directly in
the time domain\cite{Menzel}. The knowledge of this coupling is crucial to understand many
electronically induced adsorbate reactions at metal surfaces\cite{Plummer}, and it is this coupling
with the underlying substrate which governs the lifetime of the image states.

By using time-resolved 2PPE the lifetime of image-potential states on copper surfaces has been
recently determined\cite{Hertel,Wolf,Hofer2,Hofer}. Lifetimes of electrons in the $n=1$ image
state on Cu(111) have been investigated\cite{Hertel,Wolf,Hofer2}, and H\"ofer {\it et al}\cite{Hofer}
have used time-resolved 2PPE in combination with the coherent excitation of several quantum states to
investigate the ultrafast electron dynamics of image states on the (100) surface of copper,
providing an accurate experimental determination of the lifetime of the first six image states on
that surface.  

A quantitative evaluation of the lifetime of image states was first reported in
Ref.\onlinecite{Echenique3}, within a many-body free-electron description of the metal surface and
neglecting, therefore, any effects of the ion cores. Hydrogenic-like states with no penetration into
the solid were used to describe the image-state wave functions, a step model potential was
introduced to calculate the bulk final-state wave functions, simplified jellium models were used to
approximate the screened Coulomb interaction, and the image-state lifetimes were evaluated, within
the GW approximation of many-body theory\cite{Hedin}, from the knowledge of the electron
self-energy. More realistic image-state wave functions were introduced in subsequent
calculations\cite{Echenique4}, allowing for penetration into the crystal.

In this letter we present the results of a calculation of the lifetimes of the first three image
states on a Cu(100) surface and the first image state on Cu(111), by going beyond a free-electron
description of the metal surface, and including, therefore, band structure effects. First, we
evaluate the electronic wave functions by solving the time-independent Schr\"odinger equation with a
realistic one-dimensional model potential, we then use these wave functions to evaluate the screened
Coulomb interaction within a well-defined many-body framework, the random-phase approximation
(RPA)\cite{Pines}, and we finally evaluate the lifetimes from the knowledge of the imaginary part of
the electron self-energy of the excited quasiparticle, which we compute within the so-called GW
approximation.

As the image-state wave functions lie mainly in the vacuum side of the metal surface, and the
electron moves, therefore, in a region with little potential variation parallel to the
surface\cite{Hulbert}, we assume translational invariance in the plane of the surface, which is taken
to be normal to the
$z$ axis, and evaluate the damping rate of an electron in the state $\phi_0(z){\rm e}^{{\rm
i}{\bf k}_\parallel\cdot{\bf r}_\parallel}$ with energy
$E_0=(E_0)_z+{\bf k}_\parallel^2/2$ (we use atomic units throughout, i.e., $e^2=\hbar=m_e=1$), from
the knowledge of the two-dimensional Fourier transform of the electron self-energy
$\Sigma(z,z';{\bf k}_\parallel,E_{0})$, as follows
\begin{equation}
\tau^{-1}=-2\int{\rm d}{z}\int{\rm d}{z'}\phi_{0}^*(z){\rm Im}\Sigma(z,z';{\bf k}_\parallel,E_{0})
\phi_{0}(z').
\end{equation}

In the GW approximation, only the first term of the expansion of the self-energy in the screened
interaction is considered, and after replacing the Green function ($G$) by the zero order
approximation ($G^0$), one finds:
\begin{eqnarray}
{\rm Im}\Sigma(z,z';&&{\bf k}_\parallel,E_0)=\sum_{E_F\le E_f\le
E_0}\int{{\rm d}^2{\bf q}_\parallel\over(2\pi)^2}\phi_f^*(z')\cr\cr &&\times{\rm Im} W^{\rm
ind}(z,z',{\bf q}_\parallel,E_0-E_f)\phi_f(z),
\end{eqnarray}
where the sum is extended over a complete set of final states
$\phi_{f}(z){\rm e}^{{\rm i}({\bf k}_\parallel+{\bf q}_\parallel)\cdot{\bf r}_\parallel}$ with energy
$E_f=(E_f)_z+({\bf k}_\parallel+{\bf q}_\parallel)^2/2$, $E_F$ is the Fermi energy, and $W^{\rm
ind}(z,z';{\bf q}_\parallel,E)$ is the induced interaction:
\begin{eqnarray}
W^{\rm ind}(z,z';{\bf q}_\parallel,E)=\int&&{\rm d}{z_1}\int{\rm d}{z_2}
V(z,z_1;{\bf q}_\parallel)\cr
\times&&\chi(z_1,z_2;{\bf q}_\parallel,E)V(z_2,z';{\bf q}_\parallel).
\end{eqnarray}
Here, $V(z,z';{\bf q}_\parallel)$ and $\chi(z,z';{\bf q}_\parallel,E)$ represent two-dimensional
Fourier transforms of the bare Coulomb potential and the density-response function, respectively.

In the RPA, $\chi(z,z';{\bf q}_\parallel,E)$ satisfies the integral equation
\begin{eqnarray}
\chi(z,z';&&{\bf q}_\parallel,E)=\chi^0(z,z';{\bf q}_\parallel,E)
+\int{\rm d} z_1\int{\rm d} z_2\cr\cr
&&\times\chi^0(z,z_1;{\bf q}_\parallel,E)V(z_1,z_2;{\bf q}_\parallel)
\chi(z_2,z';{\bf q}_\parallel,E),
\end{eqnarray}
where $\chi^0(z_1,z_2;{\bf q}_\parallel,E)$ represents the density-response function for
non-interacting electrons. An explicit expression for $\chi^0(z_1,z_2;{\bf q}_\parallel,E)$ can be
found in Ref.\onlinecite{Eguiluz}, in terms of the eigenfunctions $\phi_i(z)$ of the one-electron
effective hamiltonian.

We compute the image-state wave function, $\phi_0(z)$, and both the final states $\phi_f(z)$ and
all the one-electron eigenfunctions involved in the evaluation of the polarizability
$\chi^0(z_1,z_2;{\bf q}_\parallel,E)$, by solving the Schr\"odinger equation with the
one-dimensional model potential recently suggested in Ref.\onlinecite{Chulkov}, which approaches, far
outside the surface, the classical image potential and describes, inside the crystal, the
self-consistent effective potential of density-functional theory. This one-dimensional potential
reproduces the width and position of the energy gap at the $\bar\Gamma$ point (${\bf
k}_\parallel=0$) and, also, the binding energies of both the
$n=0$ crystal induced surface state at $\bar\Gamma$ and the first ($n=1$) image-potential induced
state. This model potential reproduces first-principles calculations of the wave functions and
binding energies of image states in Li(100)\cite{Chulkov}, and also the average probability-density
of the
$n=1$ image state on Cu(100) derived by Hulbert {\it et al}\cite{Hulbert} from a first-principles
calculation (see Fig. 1). We note
that the members of the Rydberg series on Cu(100) with quantum number $n\le 3$  have binding energies
of
$0.57$,
$0.18$ and $0.08{\rm eV}$. The $n=1$ probability-density has a maximum at
$3.8{\rm\AA}$ outside the crystal edge ($z=0$), which we choose to be located half a lattice spacing
beyond the last atomic layer. For the fraction of the first image state overlap probability with
the bulk ($z<0$) crystal, we find $p=0.05$. On the Cu(100) surface the first image state
is closer to the center of the band gap than on the Cu(111) surface. As a result, on the latter the
penetration of the first image state into the crystal is larger, $p=0.22$, and the
probability-density has a maximum closer from the surface, at $2.3{\rm\AA}$ outside the crystal
edge. The binding energy of this image state is $0.83{\rm eV}$.

First of all, we focus on the evaluation of the damping rate of the $n=1$ image
state of Cu(100), and we set the wave vector of the image electron parallel to the surface,
${\bf k}_\parallel$, equal to zero. Coupling of the image state with the crystal occurs through the
penetration of the image-state wave function into the solid and, also, through the evanescent tails
of bulk states outside of the crystal. Accordingly, we have calculated separately the various
contributions to the damping rate by confining the integral in Eq. (1) to either bulk ($z<0$) or
vacuum ($z>0$) coordinates, and we have obtained the results presented in Table I. The decaying
rates in the bulk are expected to be larger than in the vacuum. Also, for final-state wave vectors
parallel to the surface, ${\bf q}_\parallel$, that are smaller than $\sqrt{2(E_0-E_g)}$ ($E_g$
represents the lower edge of the projected energy gap at the $\bar\Gamma$ point) there is a
reduction in the phase-space because of the presence of the projected band gap, and this results in
a reduced decay rate in the vacuum for which vertical transitions (${\bf q}_\parallel\approx 0)$ are
expected to dominate. Thus, the coupling of image states with the crystal states occurring through
the bulk penetration plays an important role in the determination of the damping rate, though the
penetration of the image-state wave function is small. The contribution to the damping rate coming
from the interference between
$z\,_{<}^{>}0$ and
$z'\,_{>}^{<}0$ coordinates is comparable in magnitude and opposite in sign to both vacuum and bulk
contributions, this being a consequence of the behaviour of the
imaginary part of the two-dimensional Fourier transform of the self-energy\cite{Eguiluz2}. The
experimentally determined damping rate reported by H\"ofer {\it et al}\,\cite{Hofer} for the first
image state on Cu(100), also presented in Table I, shows a reasonable agreement with our theoretical
prediction, though our predicted linewidth is a little larger.

Also exhibited in Fig. 1 is the square of the parametrized
hydrogenic-like wave function $\phi(z)=(4\alpha^3)^{1/2}z{\rm e}^{-\alpha z}$, with the
$\alpha$ parameter chosen so as to fit the peak position of our calculated first image-state
probability-density of Cu(100). This hydrogenic-like wave function is found to be less localized
around the maximum outside the surface, and replacement of our calculated image state wave function
by this approximation leads, therefore, to a damping rate (see Table I) that is too large. First, the
use of this approximated initial wave function gives rise to a spurious contribution from a region
just outside the surface, in which the decaying rates are noticeably larger than around the
maximum, and we note that the damping rate is highly sensitive to the actual shape of the
image-state wave function. Secondly, both the bulk contribution and the contribution from the
interference between
$z\,_{<}^{>}0$ and $z'\,_{>}^{<}0$ coordinates, which we have found to be of crucial importance, are
completely neglected within this model.

In order to investigate the impact of band structure effects on the damping rate of image
states, we have also performed a calculation in which the realistic final states
$\phi_f(z)$ considered above are replaced by the self-consistent {\it jellium} LDA one-electron wave
functions of Lang and Kohn\cite{Lang}, with and without the restriction that only final states with
energy $(E_f)_z$ lying below the projected band gap are allowed. Both the image-state wave
function and all the one-electron eigenfunctions involved in the evaluation of the screened
interaction are still obtained by introducing the one-dimensional potential of
Ref.\onlinecite{Chulkov} into the effective hamiltonian. The various contributions to the
damping rate calculated in this way are presented in Table I. For
$q_\parallel>\sqrt{2(E_0-E_g)}$ all final states with energy $E_f<E_0$ lie below the projected band
gap; thus, the bulk and interference contributions to the damping rate remain almost unaffected by
this restriction. However, as the coupling of the image state with the crystal occurring through
the tails of bulk states outside of the crystal is expected to be dominated by vertical
transitions (${\bf q}_\parallel\approx 0$), the vacuum contribution to the damping rate becomes
dramatically smaller as final states lying within the projected band gap are not allowed.

As far as the screened
interaction is concerned, systematic investigations of the role that
this quantity plays in the coupling of image states with the solid  have been
performed\cite{Sarria}. We find that simplified jellium models for the electronic response
lead to unrealistic results for the lifetime, though the impact of the band structure in the
evaluation of the screened interaction is not large. 

In a heuristic view, the bulk contribution to the decay rate for an image state might be
approximated by the value of the decay rate for a bulk state at the same energy, times the fraction
of the image state overlap probability with the bulk crystal. We have calculated, within the GW-RPA,
the decay rate for bulk states in a homogeneous electron gas, thereby neglecting band structure
effects in both the initial $\phi_0(z)$ and final states $\phi_f(z)$ and, also, in the
polarizability. The damping rate for a bulk state at the energy of the first image state on Cu(100),
times the penetration of this image state, results in a linewidth of $14{\rm meV}$\cite{note1}, well
below the bulk contribution to our full GW-RPA surface calculation presented in Table I.

Finally, we consider the lifetime, $\tau$, of image-potential states on Cu(100) with quantum
number $n\le 3$, and the first image state on Cu(111). As before,
image-state wave functions, final states and all the eigenfunctions involved in the evaluation of the
polarizability are obtained by solving the Schr\"odinger equation with the one-dimensional model
potential of Ref.\onlinecite{Chulkov}, and we set ${\bf k}_\parallel=0$. The results of these
calculations are presented in Table II, together with the experimentally determined decay rates
reported in Refs.\onlinecite{Wolf,Hofer2} and \onlinecite{Hofer} for Cu(111) and Cu(100),
respectively. We observe that both our calculated and the experimentally determined lifetimes of the
image states on Cu(100) increase rapidly with
$n$ as $\sim 1/n^3$, in agreement with previous theoretical
predictions\cite{Echenique2}; this is a result of the weaker spatial overlap with the bulk
that higher order image states present. As for the lifetime of the first image state
on Cu(111) we find that it is comparable to that of Cu(100), though this image
state on Cu(111) is located close to the top of the projected band gap and the wave function
overlap with the bulk is, therefore, much larger. Also, the surface electronic structure of
Cu(111) supports the well-known $n=0$ crystal-induced surface state, and we find that the decaying
rate of the $n=1$ image state to this intrinsic surface state results in a linewidth of $16{\rm
meV}$, which represents a $40\%$ of the total linewidth ($\tau^{-1}=38{\rm meV}$). However, both
the large bulk-state overlap and the existence of the $n=0$ surface state are counterbalanced by
the band gap extending on the Cu(111) surface below the Fermi level and the available phase space
becoming, therefore, highly restricted.

In conclusion, we have reported theoretical calculations of the lifetime of image-potential states
on copper surfaces, by going beyond a free-electron description of the metal surface. We have analyzed the origin and magnitude of the
various contributions to the quasiparticle damping. Our results indicate that contributions to the
damping rate coming from the bulk and from the interference between bulk and vacuum coordinates
both play an important role in the determination of the lifetimes. We have also demonstrated that
the vacuum contribution to the damping rate is highly sensitive to the location of the projected
band gap. Our results are in reasonably good agreement with the experimentally determined decay times
reported by H\"ofer {\it et al}\cite{Hofer} in the case of Cu(100), and our calculated lifetime of
the first image state on Cu(111) is in excellent agreement with the experimental results reported in
Refs.\onlinecite{Wolf} and \onlinecite{Hofer2}. 

We acknowledge partial support by the Basque Hezkuntza, Unibertsitate eta Ikerketa Saila, the Spanish
Ministerio de Educaci\'on y Cultura, and Iberdrola S.A.

\begin{table}
\caption{Damping rates, in linewidth units (meV), of the $n=1$ image state on Cu(100). (a) Our full
calculation, as described in the text, together with the experimental damping rate
of Ref.\protect\onlinecite{Hofer}\protect. (b) The result of our calculation when our realistic
image-state wave function is replaced by the hydrogenic-like wave function described in the text.
(c) and (d) The result of our calculation when our realistic final states are replaced by
the self-consistent jellium LDA wave functions of Lang and Kohn\protect\cite{Lang}\protect, (c) with
and (d) without the restriction that only final states with energy \protect$(E_f)_z$\protect\, lying
below the projected band gap are allowed.} 
\begin{tabular}{lcccccccc}
&Bulk&Vacuum&Interference
&Total&Experiment\\
\tableline
(a)&24&14&-16&22&16.5\\
(b)&-&71&-&71\\
(c)&19.5&13&-11&21.5\\
(d)&21&58&-12&67\\
\end{tabular}
\label{table1}
\end{table}

\begin{table}
\caption{Calculated lifetimes, in femtoseconds, of the \protect$n\le 3$\protect\, image states on
Cu(100) and the first image state on Cu(111), together with the experimentally determined lifetimes
of Refs.\protect\onlinecite{Wolf,Hofer2}\protect\, and \protect\onlinecite{Hofer}\protect\, for
Cu(111) and Cu(100), respectively. The damping rate of Eq. (1) is related to the
lifetime,
\protect$\tau$\protect, by
\protect$\tau^{-1}\tau=1{\rm a.u.}=660\,{\rm meV}\,{\rm fs}$\protect.} 
\begin{tabular}{lcccccccc}
&\multicolumn{2}{c}{Cu(100)}&\multicolumn{4}{c}{Cu(111)}\\
&Theory&Experiment\protect\cite{Hofer}\protect&Theory&Experiment\protect\cite{Wolf,Hofer2}\protect
&&\\
\tableline
$n=1$&30&$40\pm 6$&17.5&$18\pm 5$\protect\cite{Wolf}\protect&\\
     &  &         &    &$15\pm 5$\protect\cite{Hofer2}\protect&\\
$n=2$&132&$110\pm 10$&&\\
$n=3$&367&$300\pm 15$&\\
\end{tabular}
\label{table2}
\end{table}

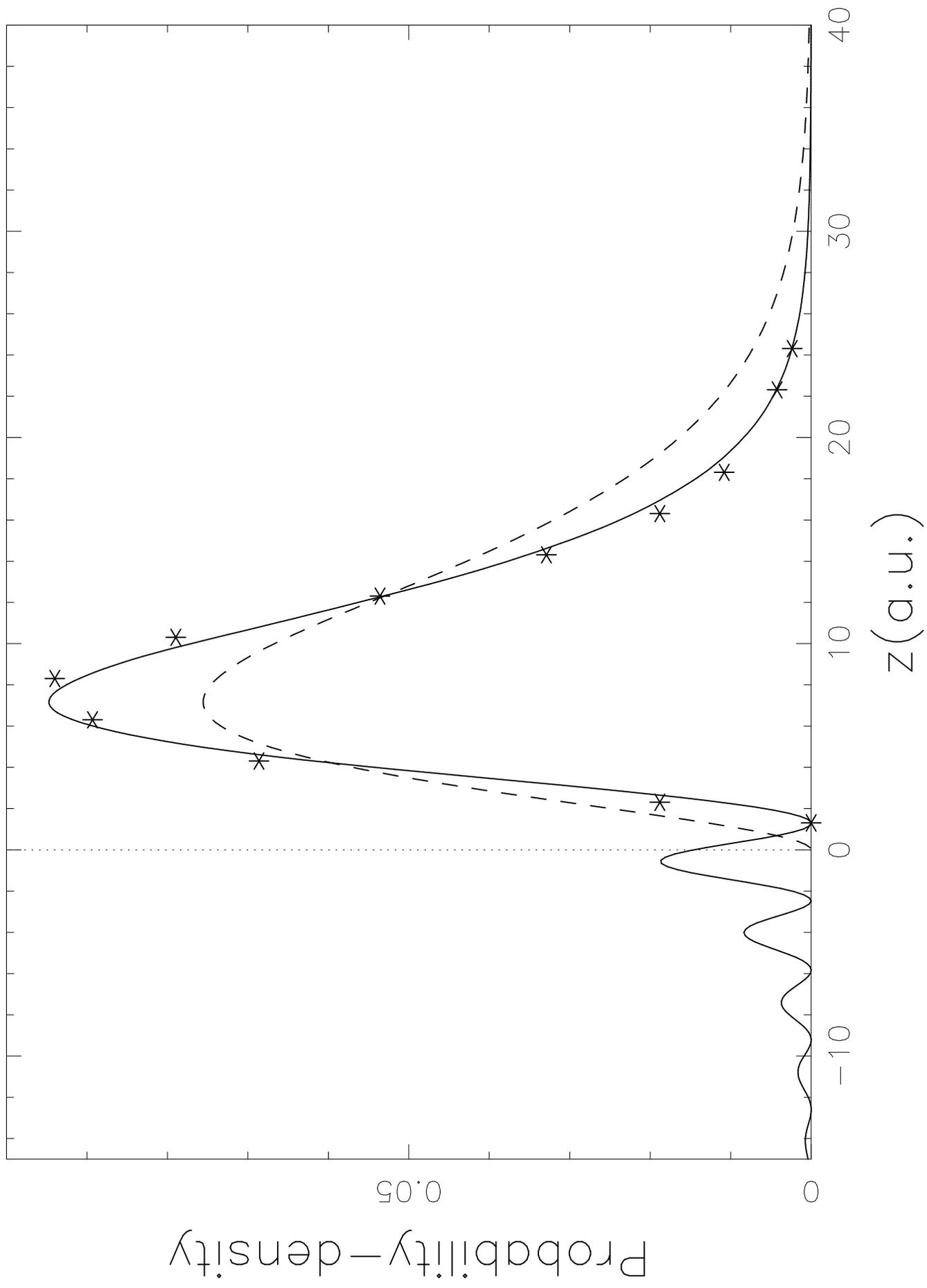
\begin{figure}
\caption{Probability-density of the $n=1$ image state on Cu(100) (solid line) and
the hydrogenic-like approximation described in the text (dashed line). Stars represent the
probability-density of the $n=1$ image state, averaged parallel to the surface, as reported in
Ref.\protect\onlinecite{Hulbert}\protect.}
\end{figure}

\end{document}